\documentclass[a4paper]{jpconf}
\usepackage{graphicx}
\begin{document}
\title{A data-driven MHD model of the global solar corona within Multi-Scale Fluid-Kinetic Simulation Suite (MS-FLUKSS)} 

\author{Mehmet Sarp Yalim$^2$, Nikolai Pogorelov$^{1,2}$, and Yang Liu$^3$}

\address{$^1$Department of Space Science, The University of Alabama in Huntsville, 320 Sparkman Drive, Huntsville, AL 35805, USA}
\address{$^2$Center for Space Plasma and Aeronomic Research, The University of Alabama in Huntsville, 320 Sparkman Drive, Huntsville, AL 35805, USA}
\address{$^3$W. W. Hansen Experimental Physics Laboratory, Stanford University, 452 Lomita Mall, Stanford, CA 94305, USA}

\ead{msy0002@uah.edu,np0002@uah.edu,yliu@sun.stanford.edu}

\begin{abstract}
We have developed a data-driven magnetohydrodynamic (MHD) model of the global solar corona which uses characteristically-consistent boundary conditions (BCs) at the inner boundary. Our global solar corona model can be driven by different observational data including Solar Dynamics Observatory/Helioseismic and Magnetic Imager (\emph{SDO/HMI}) synoptic vector magnetograms together with the horizontal velocity data in the photosphere obtained by the time-distance helioseismology method, and the line-of-sight (LOS) magnetogram data obtained by \emph{HMI}, Solar and Heliospheric Observatory/Michelson Doppler Imager (\emph{SOHO/MDI}), National Solar Observatory/Global Oscillation Network Group (\emph{NSO/GONG}) and Wilcox Solar Observatory (\emph{WSO}). We implemented our model in the Multi-Scale Fluid-Kinetic Simulation Suite (MS-FLUKSS) -- a suite of adaptive mesh refinement (AMR) codes built upon the Chombo AMR framework developed at the Lawrence Berkeley National Laboratory. We present an overview of our model, characteristic BCs, and two results we obtained using our model: A benchmark test of relaxation of a dipole field using characteristic BCs, and relaxation of an initial PFSS field driven by HMI LOS magnetogram data, and horizontal velocity data obtained by the time-distance helioseismology method using a set of non-characteristic BCs.   

\end{abstract}

\section{Introduction}
\label{sec1}

The solar wind (SW) emerging from the Sun is the main driving mechanism of solar events which may lead to geomagnetic storms which are the primary causes of space weather disturbances that affect the magnetic environment of Earth and may have hazardous effects on the space-borne and ground-based technological systems as well as human health. Therefore, accurate modeling of the SW is very important to understand the underlying mechanisms of such storms.

As the number of space-borne and ground-based observatories increased significantly over the past decades, a number of data-driven models of SW that involve formulations of time-dependent boundary conditions (BCs) that incorporate remote and \textit{in situ} observations of the Sun in a self-consistent way have been developed (see, e.g. ~\cite{Mikic99,Wu06,Hayashi05,vanderHolst10,Yang12,Feng12}). 

The SW emerges from the Sun. In order to model the background SW in a physically-consistent way, it is essential to apply a proper boundary treatment based on observational data on the Sun. For this purpose, we have developed a data-driven magnetohydrodynamic (MHD) model of the global solar corona which applies characteristically-consistent BCs at the inner boundary of our computational domain located in the lower solar corona. Our characteristic BC formulation follows~\cite{Hirsch94}, is \textit{entirely based on observations, and does not rely on any non-reflection principle, as in}~\cite{Thompson87}. It allows us to specify sufficient number of \textit{mathematically admissible} BCs. The implementation of characteristic BCs at the lower corona is based not only on the knowledge of how many quantities should be specified, but also on the requirement that time-increments of quantities that are specified as physical BCs should be uniquely expressible in terms of the increments of proper characteristic variables. 

With the availability of vector magnetograms and surface velocity components, we can impose our characteristic BC formulation to obtain a realistic time-dependent background SW model. Our model is driven by Solar Dynamics Observatory/Helioseismic and Magnetic Imager (\emph{SDO/HMI}) synoptic vector magnetograms together with the horizontal velocity data in the photosphere obtained, e.g., with the time-distance helioseismology method~\cite{Zhao12}. In addition, our model can also be driven by other observational data including the line-of-sight (LOS) magnetogram data obtained by \emph{HMI}, Solar and Heliospheric Observatory/Michelson Doppler Imager (\emph{SOHO/MDI}), National Solar Observatory/Global Oscillation Network Group (\emph{NSO/GONG}), and Wilcox Solar Observatory (\emph{WSO}).   

We obtain the initial distribution of plasma properties in the computational domain from the Potential Field Source Surface (PFSS) model~\cite{AN69,Schatten69} for the magnetic field and Parker's 1D isothermal SW model~\cite{Parker58} for the hydrodynamic variables. We model the SW acceleration by introducing volumetric heating source terms into the momentum and energy equations that incorporate expansion factors to take the coronal magnetic field distribution into account~\cite{Nakamizo09,Feng10}.

We have implemented our model in the Multi-Scale Fluid-Kinetic Simulation Suite (MS-FLUKSS)~\cite{Pogorelov14} -- a suite of adaptive mesh refinement (AMR) codes designed to solve the coupled systems of MHD, gas dynamics Euler, and kinetic Boltzmann equations~\cite{Borovikov09,Pogorelov09,Borovikov13,Pogorelov13}. MS-FLUKSS is built upon the Chombo AMR framework developed at the Lawrence Berkeley National Laboratory. MS-FLUKSS also has modules that treat pickup ions as a separate fluid and turbulence beyond the Alfv\'enic surface.

In section~\ref{sec2}, we will present an overview of our global solar corona model with particular emphasis given to our characteristic BC formulation and \emph{SDO/HMI} vector magnetogram and horizontal velocity data. In section~\ref{sec3}, we will present results we obtained using our model. Finally, in section~\ref{sec4}, we will present our conclusions.  

\section{Data-driven MHD model of global solar corona}
\label{sec2}

\subsection{Governing equations}
\label{sec2_1}
We solve the set of ideal MHD equations with volumetric heating source terms to model acceleration of SW~\cite{Nakamizo09,Feng10}. They are written in terms of conservative variables, in conservation-law form as follows:

\begin{equation}
\label{eq1}
\frac{\partial}{\partial t}
\left(
\begin{array}{c}
\rho  \\
\rho \mathbf{v} \\
\mathbf{B} \\
E
\end{array}
\right) +
\mathbf{\nabla}\cdot
\left(
\begin{array}{c}
\rho\mathbf{v} \\
\rho\mathbf{v}\mathbf{v} + \mathbf{I}(p+\frac{B^2}{8\pi})-\frac{\mathbf{B}\mathbf{B}}{4\pi} \\
\mathbf{v}\mathbf{B}-\mathbf{B}\mathbf{v} \\
(E+p+\frac{B^2}{8\pi})\mathbf{v}-\frac{\mathbf{B}}{4\pi}(\mathbf{v}\cdot\mathbf{B})
\end{array}
\right) =
\left(
\begin{array}{c}
0 \\
\rho[\mathbf{g}+(\mathbf{\Omega}\times\mathbf{r})\times\mathbf{\Omega}+2(\mathbf{v}\times\mathbf{\Omega})] + \mathbf{{S}_{M}} \\
0 \\
\rho\mathbf{v}\cdot[\mathbf{g}+(\mathbf{\Omega}\times\mathbf{r})\times\mathbf{\Omega}] + S_{E}
\end{array}
\right),
\end{equation} where $\rho$, $\mathbf{v}$, $\mathbf{B}$, $p$, $E$, and $\mathbf{g}$ are the density, velocity, magnetic field, thermal pressure, specific total energy of the plasma, and gravitational acceleration, respectively. The source terms in the momentum and energy conservation equations include the Coriolis and centrifugal forces which are present when the system is solved in a frame corotating with the Sun. Accordingly, $\Omega$ and $\mathbf{r}$ correspond to the angular velocity of the Sun and position vector, respectively.

In order to model the acceleration of the SW, we introduce a volumetric heating source term, $S_{E}$, into the energy conservation equation, and the corresponding source term, $\mathbf{{S}_{M}}$, into the conservation of momentum equations~\cite{Nakamizo09,Feng10}. They are given as follows:

\begin{equation}
\label{eq2a}
S_{E}=\frac{Q_{0}}{f_{s}}\textrm{exp}\Big(-\frac{r}{L_{Q}}\Big), \\
\end{equation}

\begin{equation}
\label{eq2b}
\mathbf{S_{M}}=\frac{M_{0}}{f_{s}}\Big(\frac{r}{R_{\odot}}-1\Big)\textrm{exp}\Big(-\frac{r}{L_{M}}\Big),
\end{equation} where $L_\mathrm{M}$, $L_\mathrm{Q}$, $M_0$, and $Q_0$ are the model constants given as $L_\mathrm{M}=L_Q=0.9 R_\odot$, $M_0=2.65\times 10^{-14}$ N $\mathrm{m}^{-3}$, and $Q_0=1.65\times 10^{-6}$ J $\mathrm{m}^{-3}$$\mathrm{s}^{-1}$. Additionally, $f_{s}$ is the expansion factor by which a magnetic flux tube expands in solid angle between its footpoint location on the photosphere and the source surface which is typically at $R_\mathrm{SS}=2.5R_\odot$~\cite{WS97}:

\begin{equation}
\label{eq3}
f_{s}=\frac{B(R_\odot)}{B(R_{SS})}\left(\frac{R_\odot}{R_{SS}}\right)^2.
\end{equation} 

These source terms take the coronal magnetic field topology into account by incorporating the expansion factor. 

The expansion factor is initially computed in every cell located between the inner boundary and the source surface and kept constant throughout a simulation. 
\subsection{Observational data}
\label{sec2_2}
Our model is primarily driven by full disk \emph{HMI} synoptic vector magnetograms~\cite{Liu17} and horizontal velocity data in the photosphere.

\emph{HMI}~\cite{Schou12} provides full disk vector magnetogram data with high cadence ($\sim$ 720 s) and no data gaps~\cite{Hoeksema14}. From these data, synoptic vector magnetograms are computed for each Carrington rotation for the entire \emph{SDO}~\cite{Pesnell12} mission. They have a resolution of 3600$\times$1440 pixels and are provided by the \textit{Joint Science Operations Center} (JSOC) under the data series name hmi.B\_synoptic~\cite{Liu17}. 

The horizontal velocity data are inferred by the time-distance helioseismology method~\cite{Zhao12}. The acoustic travel times are measured using \emph{HMI} Dopplergram observations under the data series name hmi.V and this method infers solar interior properties by inverting these measurements. Thus, near-real time full disk maps of subsurface wave-speed perturbations and horizontal flow velocities are produced for depths ranging from 0 to 20 \textrm{Mm}, every 8 hours. Carrington synoptic maps for the subsurface properties are made from these full disk maps. These maps have a resolution of 3000$\times$1000$\times$6 pixels where the last dimension corresponds to the depth beneath the solar surface. We utilize the corresponding data on the photosphere.

\begin{figure}[!h]
\centering
\includegraphics[height=5.5cm]{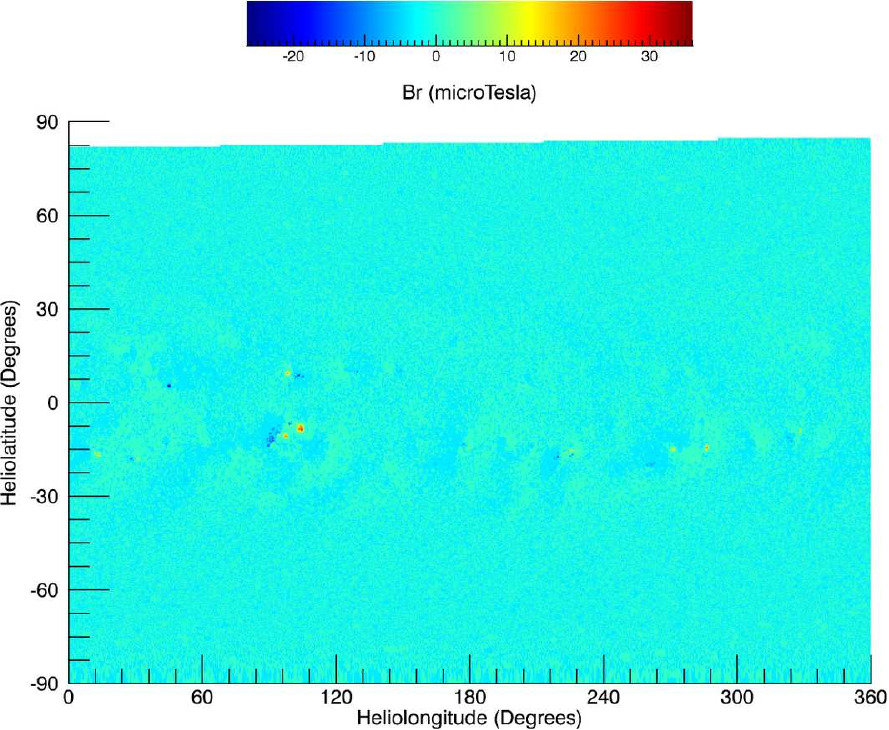} 
\includegraphics[height=5.5cm]{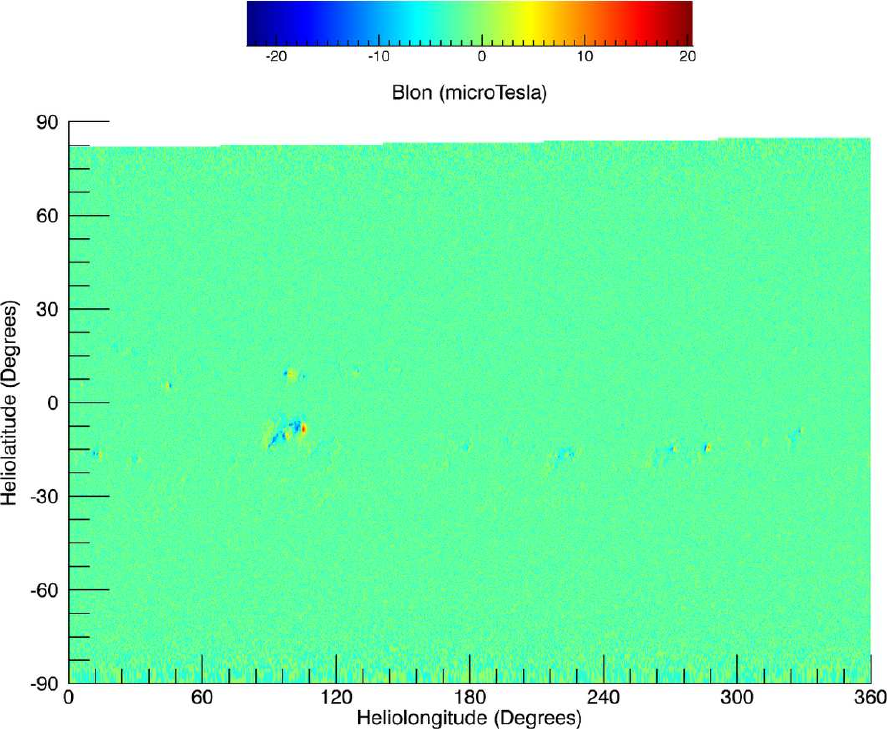}\\ 
\begin{center}
\includegraphics[height=5.5cm]{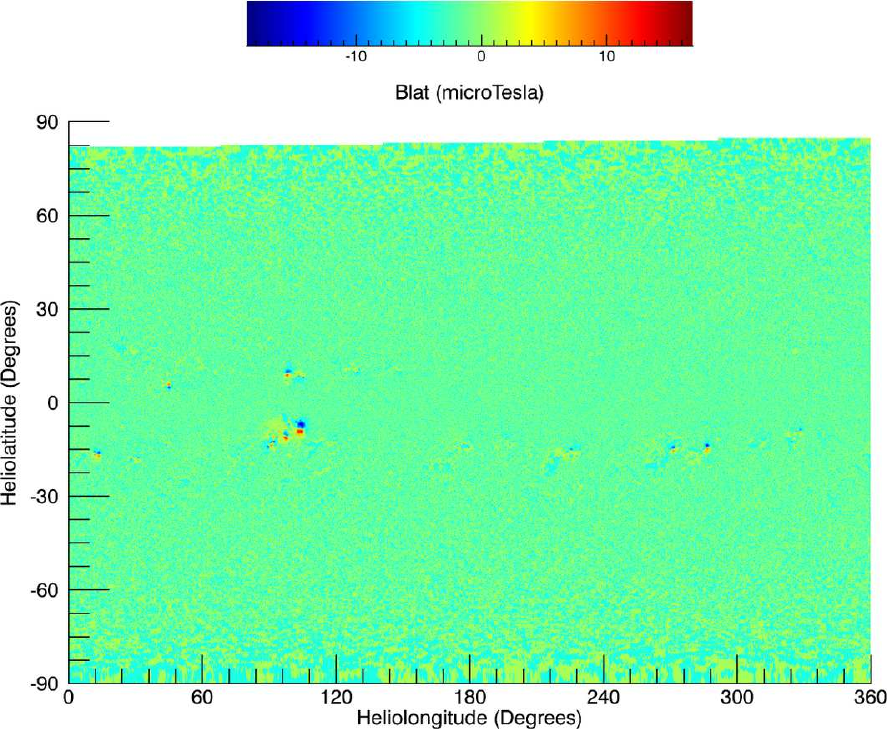}
\end{center}
\caption{\emph{HMI} vector field synoptic chart for CR 2145: (\textit{Left panel}) radial, (\textit{Right panel}) longitudinal, and (\textit{Bottom panel}) latitudinal components.}
\label{fig1}
\end{figure}

\begin{figure}[!h]
\centering\hfill\includegraphics[height=5.5cm]{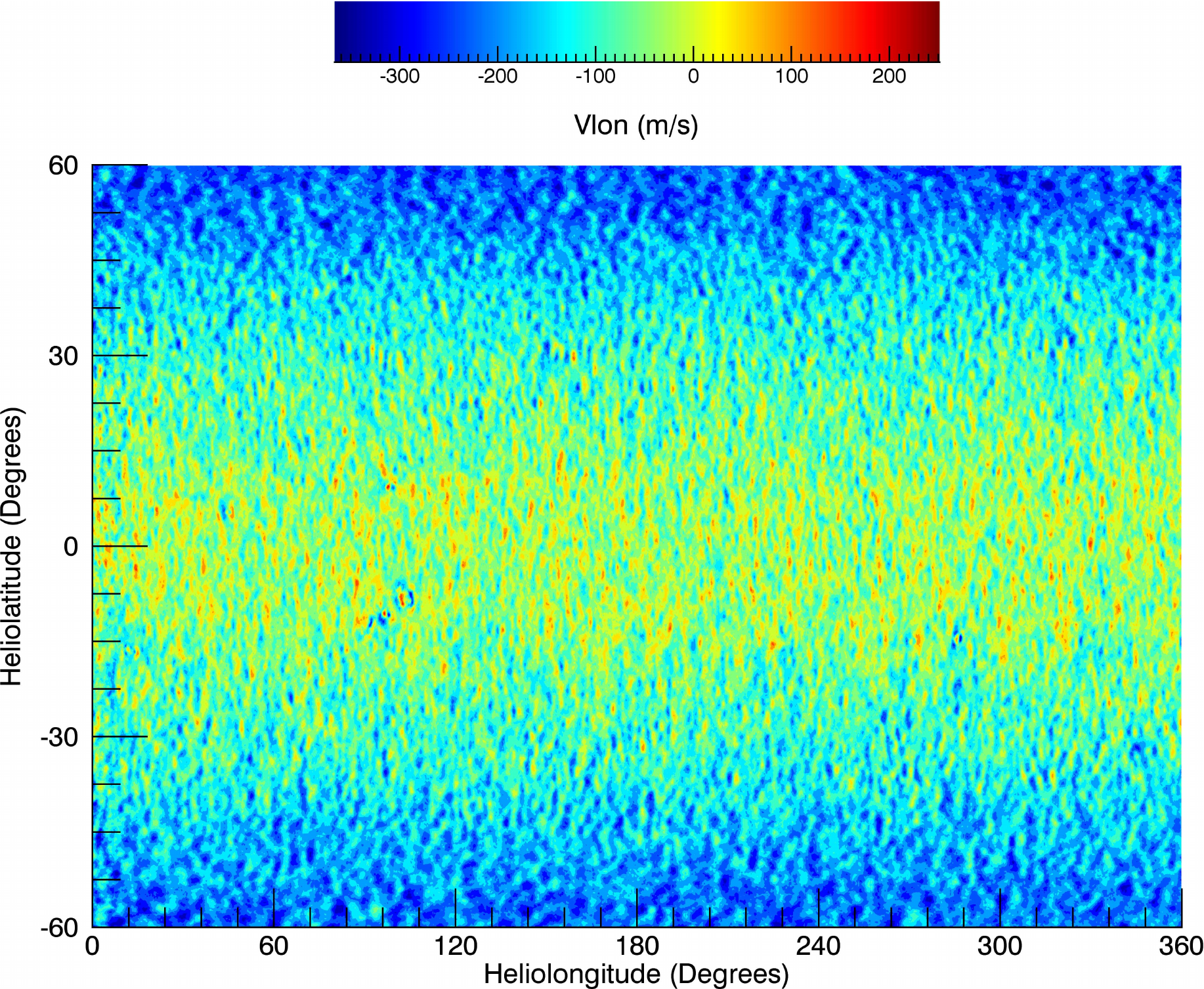}\hfill
\includegraphics[height=5.5cm]{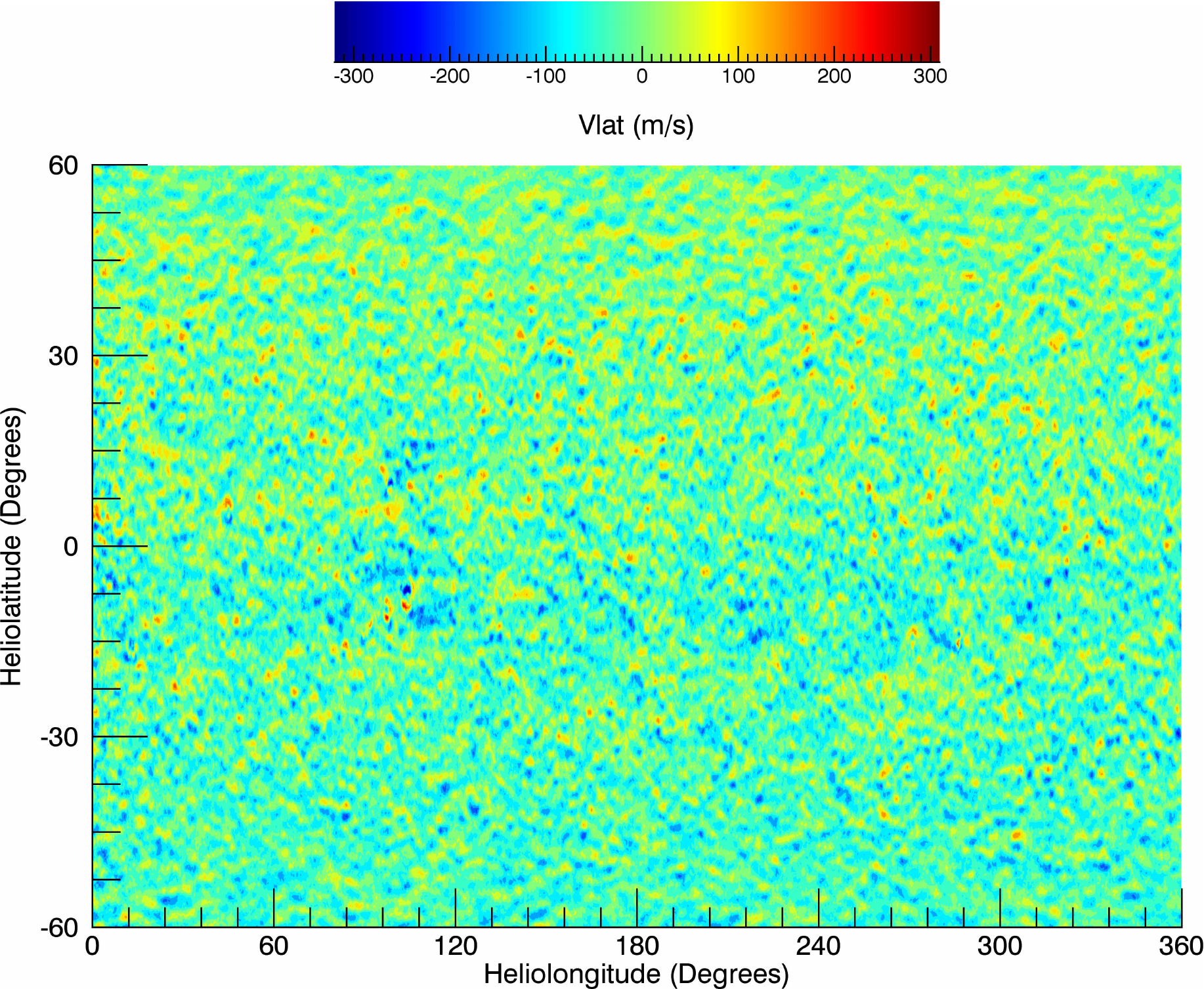}\hspace*{\fill}
\caption{Synoptic chart of the horizontal velocity data for CR 2145: (\textit{Left panel}) longitudinal, and (\textit{Right panel}) poleward flows.}
\label{fig2}
\end{figure}

In Figures~\ref{fig1} and~\ref{fig2}, three components of the magnetic field vector and horizontal velocity components are shown on the solar surface corresponding to Carrington Rotation (CR) 2145, respectively. The horizontal velocity profiles are given within the latitude range of $\pm 60^{\circ}$. Above this latitude, the inferred velocity is not very reliable.

Our model can also be driven by other observational data including \emph{NSO/GONG}, \emph{HMI}, \emph{SOHO/MDI}, and \emph{WSO} LOS magnetogram data.
\subsection{Initial solution}
\label{sec2_3}
We calculate the initial distribution of magnetic field using the PFSS model~\cite{AN69,Schatten69} based either on the spherical harmonics approach~\cite{Hoeksema84,WS92,SD03} or a finite difference method by incorporating the solution provided by the Finite Difference Iterative Potential-field Solver (FDIPS) code~\cite{TVH11} into MS-FLUKSS. FDIPS is a good alternative to the spherical harmonics approach. As the number of spherical harmonics increases, using raw magnetogram data specified on a grid that is uniform in the sine of latitude can result in inaccurate and unreliable results, especially in the polar regions close to the Sun. As the FDIPS code accepts any magnetogram data represented in the Flexible Image Transport System (FITS) file format as input, we can also incorporate potential magnetic field solutions for all such data into MS-FLUKSS. For the rest of the plasma parameters, we compute the \textit{initial solution} from Parker's isothermal SW model~\cite{Parker58}.
\subsection{Characteristic boundary conditions}
\label{sec2_4}
 One of the important components of our data-driven MHD model is a genuinely characteristic BC formulation, which became possible only recently because of the availability of vector magnetograms. If the inner simulation boundary is in the vicinity of the photosphere ($R\approx R_\odot$), the radial component of the SW velocity is likely to be less than the Alfv\'en or even slow magnetosonic velocity. Not only this determines the number of physical BCs at $R\approx R_\odot$, but the characteristic variables to be specified are also uniquely defined.
Since there are no analytic expressions for characteristic variables in MHD, they cannot be found from observational data.
On the other hand, any admissible set of physical BCs should be resolvable for the above-mentioned physical characteristic variables. The remaining BCs should be mathematical (numerical). They should be specified, e.g., by extrapolation of the corresponding variables from inside the computational region. This approach is widely applied in aerodynamic simulations
\cite{Hirsch94}.

Consider the system of governing equations written in terms of the so-called primitive variables (the abbreviations are conventional).
\begin{eqnarray}
\label{eq4}
\frac{\partial \rho}{\partial t}+\mathbf{\nabla}\cdot(\rho\mathbf{v})=0 \nonumber \\
\frac{\partial\mathbf{v}}{\partial t}+(\mathbf{v}\cdot\mathbf{\nabla})\mathbf{v}=-\frac{\mathbf{\nabla} p}{\rho}-\frac{\mathbf{B}\times(\mathbf{\nabla}\times\mathbf{B})}{4\pi\rho}+\mathbf{g}+\mathbf{S}_\mathrm{M} \nonumber \\
\frac{\mathrm{d} p}{\mathrm{d} t}=a^{2}\frac{\mathrm{d} \rho}{\mathrm{d} t}+S_\mathrm{E} \\
\frac{\partial\mathbf{B}}{\partial t}=\mathbf{\nabla}\times(\mathbf{v}\times\mathbf{B}) \nonumber \\
\mathbf{\nabla}\cdot\mathbf{B}=0. \nonumber
\end{eqnarray}

Assuming spherical coordinates for definiteness, we rewrite this system in the quasilinear form as follows:
\begin{equation}\label{eq5}
\frac{\partial\mathbf{u}}{\partial t}+\bar{A}_{r}\frac{\partial\mathbf{u}}{\partial r}+\bar{A}_{\theta}\frac{\partial\mathbf{u}}{\partial \theta}+\bar{A}_{\phi}\frac{\partial\mathbf{u}}{\partial \phi}=\mathbf{S},
\end{equation}
where $\mathbf{u}$ is the vector of primitive variables, $\bar{A}_{r}$, $\bar{A}_{\theta}$, and $\bar{A}_{\phi}$ are the coefficient matrices in front of the $r$-, $\theta$-, and $\phi$-derivatives, respectively, and $\mathbf{S}$ is the source term vector involving the gravitational force, spherical geometrical factors, and volumetric heating terms.

Let $\mathbf{H}=\bar{A}_{\theta}\frac{\partial\mathbf{u}}{\partial \theta}+\bar{A}_{\phi}\frac{\partial\mathbf{u}}{\partial \phi}-\mathbf{S}$. Then, Eq. (\ref{eq5}) becomes
\begin{equation}
\label{eq6}
\frac{\partial\mathbf{u}}{\partial t}+\bar{A}_{r}\frac{\partial\mathbf{u}}{\partial r}+\mathbf{H}=0.
\end{equation}

To write out a set of compatibility relations, we multiply the above equation by the left eigenvector matrix of $\bar{A}_{r}$~\cite{Thompson87,Thompson90}:
\begin{equation}
\label{eq7}
\bar{L}\frac{\partial\mathbf{u}}{\partial t}+\bar{L}\bar{A}_{r}\frac{\partial\mathbf{u}}{\partial r}+\bar{L}\mathbf{H}=0,
\end{equation}
or in the component form (note that $\mathbf{l}_{i}\bar{A}_{r}=\lambda_{i}\mathbf{l}_{i}$)
\begin{equation}
\label{eq8}
\mathbf{l}_{i}\cdot\frac{\partial\mathbf{u}}{\partial t}+\lambda_{i}\mathbf{l}_{i}\cdot\frac{\partial\mathbf{u}}{\partial r}+\mathbf{l}_{i}\cdot\mathbf{H}=0,
\end{equation}
where $\mathbf{l}_{i}$ and $\lambda_{i}$ are the rows of the left eigenvector matrix $\bar{L}$ and eigenvalues of $\bar{A}_{r}$, respectively.

On introducing a set of characteristic variables $\mathbf{W}$, such that $\delta\mathbf{W}=\bar{L}\delta\mathbf{u}$, we obtain a diagonal system ($\Lambda = \{\lambda_1, \lambda_2,\ldots, \lambda_8\}$)
\begin{equation}
\label{eq9}
\frac{\partial\mathbf{W}}{\partial t}+\Lambda\frac{\partial\mathbf{W}}{\partial r}+\bar{L}\mathbf{H}=0.
\end{equation}
No analytic expressions for $\mathbf{W}$ exist in MHD, but their increments are well defined.

At the inner (SW entrance) boundary, $W_{i}$ ($i=1,\ldots k$) corresponding to $\lambda_{i}>0$ (incoming characteristics) should be specified as physical BCs, whereas the BCs for remaining characteristic variables should be numerical (i.e., they should be derived from the characteristic compatibility relations for ${W}_{j}$ with $\lambda_{j}<0$ for $j=k+1,\ldots n$). For MHD equations, $n=8$.

Following this mathematical principle, we split the compatibility relations (i.e., Eq. (\ref{eq9})) into two sets corresponding to the physical (\textit{P}) and numerical (\textit{N}) BCs:

\begin{equation}
\label{eq10}
\frac{\partial}{\partial t}
\left|
\begin{array}{c}
\mathbf{W}^{P} \\
\mathbf{W}^{N}
\end{array}
\right| +
\left|
\begin{array}{c}
\bar{L}^{P} \\
\bar{L}^{N}
\end{array}
\right|
\bar{A}_{r}
\frac{\partial}{\partial r}
\left|
\begin{array}{c}
\mathbf{u}^{P} \\
\mathbf{u}^{N}
\end{array}
\right| +
\bar{L}\mathbf{H}=0,
\end{equation}
where $\mathbf{W}^{P}$, $\mathbf{W}^{N}$, $\bar{L}^{P}$, and $\bar{L}^{N}$ are comprised of the characteristic variables and left eigenvectors corresponding to incoming and outgoing characteristics, respectively, and $\mathbf{u}^{P}$ and $\mathbf{u}^{N}$ are the primitive variables to be physically imposed or numerically determined.

The physical BCs are implemented by setting 

\begin{equation}
\label{eq11}
\frac{\partial \mathbf{W}^{P}}{\partial t}
=
(\bar{L}^{P})^{-1}
\mathbf{C}^{P}(t)\textrm{ or } 
\frac{\partial \mathbf{u}^{P}}{\partial t}=\mathbf{C}^{P}(t), 
\end{equation}
where $\mathbf{C}^{P}(t)$ is the vector of time dependent variations of the primitive variables that are physically imposed. At each time step, $\mathbf{u}^{P}$ values are interpolated in time using successive observational data (e.g., vector magnetogram) with the chosen order of accuracy.

Keeping in mind Eq. (\ref{eq11}), we can rewrite Eq. (\ref{eq10}) in terms of the primitive variables: 

\begin{equation}
\label{eq12}
\left|
\begin{array}{c}
\bar{I}^{P} \\
\bar{L}^{N}
\end{array}
\right|
\frac{\partial}{\partial t}
\left|
\begin{array}{c}
\mathbf{u}^{P} \\
\mathbf{u}^{N}
\end{array}
\right| +
\left|
\begin{array}{c}
0 \\
\bar{L}^{N}
\end{array}
\right|
\bar{A}_{r}
\frac{\partial}{\partial r}
\left|
\begin{array}{c}
\mathbf{u}^{P} \\
\mathbf{u}^{N}
\end{array}
\right| +
\left|
\begin{array}{c}
-\mathbf{C} \\
\bar{L}^{N}\mathbf{H}
\end{array}
\right|
=0,
\end{equation}
where $\bar{I}^{P}$ is a matrix of size $(P\times(N+P))$ with $I^{P}_{ii}=1$ and $I^{P}_{ij}=0$ for $i\neq j$, and $\mathbf{C}=\left|\begin{array}{c}\mathbf{C}^{P}(t) \\ 0\end{array}\right|$. Let us define $\bar{L}_{1}=\left|\begin{array}{c}\bar{I}^{P} \\\bar{L}^{N}\end{array}\right|$, and $\bar{L}_{2}=\left|\begin{array}{c} 0 \\ \bar{L}^{N}\end{array}\right|$.

After making necessary substitutions, the set of primitive variables at the inner boundary can be determined from 

\begin{equation}
\label{eq13}
\frac{\partial}{\partial t}
\left|
\begin{array}{c}
\mathbf{u}^{P} \\
\mathbf{u}^{N}
\end{array}
\right| =
\mathbf{C}-(\bar{L}_{1})^{-1}\bar{L}_{2}\bar{A}_{r}\frac{\partial}{\partial r}
\left|
\begin{array}{c}
\mathbf{u}^{P} \\
\mathbf{u}^{N}
\end{array}
\right| - (\bar{L}_{1})^{-1}\bar{L}_{2}\mathbf{H}.
\end{equation}

It should be noted that our BCs are evolutionary (i.e., small changes in them lead to small changes in solution, see~\cite{KPS01}), iff the left eigenvector matrix of the split system, $\bar{L}_{1}$, is non-singular. This determines the set of admissible physical BCs in terms of the primitive variables.
It follows from our analysis that for realistic inflow condition only the plasma density and the radial component of magnetic field
should always be specified as physical BCs. The choice of other physical BCs in terms of the primitive variables depends on the
availability of appropriate measurements. Table~\ref{tab1} describes reasonable choices, as well as the number of physical and numerical BCs.

\begin{table}[h!]
\centering
\footnotesize
\begin{center}
\begin{tabular}{|p{2.8cm}|p{2.8cm}|p{2.8cm}|p{2.8cm}|p{2.8cm}|}
\hline
Case & \# of physical BCs & \# of numerical BCs & Variables to be imposed as physical BCs & Variables to be imposed as numerical BCs \\
\hline
Subslow inflow ($v_r<a_\mathrm{s})$& 5 & 3 & $\rho$, $B_{r}$, $v_{\theta}$, $v_{\phi}$, and $B_{\theta}$ & $v_{r}$, p, and $B_{\phi}$ \\
\hline
Sub-Alfv\'enic, but superslow~inflow ($a_\mathrm{s}<v_r<a_\mathrm{A}$)& 6 & 2 & $\rho$, $v_{\theta}$, $v_{\phi}$, p, $B_{r}$, and $B_{\theta}$ & $v_{r}$, and $B_{\phi}$ \\
\hline
Subfast, but super-Alfv\'enic~inflow ($a_\mathrm{A}<v_r<a_\mathrm{f}$)& 7 & 1 & $\rho$, $v_{\theta}$, $v_{\phi}$, p, $B_{r}$, $B_{\theta}$, and $B_{\phi}$ & $v_{r}$ \\
\hline
\end{tabular}
\end{center}
\caption{Different ways to specify the characteristic BCs. The slow magnetosonic, Alfv\'enic, and fast magnetosonic velocities are denoted as $a_\mathrm{s}$,
$a_\mathrm{A}$, and $a_\mathrm{f}$, respectively.}
\label{tab1}
\end{table}

Thus, it can be seen that \emph{SDO/HMI} data allow us to specify the necessary BCs, except for density.
It was demonstrated in~\cite{Lionello09} that underestimating or overestimating density at the inner boundary does not affect the coronal solution. Hence, we can assume uniform density at the inner boundary since we do not have measurements.

We discretize our characteristic compatibility equations (see Eq. (\ref{eq13})) using TVD, Roe-type schemes. For the components of the vector magnetogram and horizontal velocity data which can be imposed as physical BCs (see Table~\ref{tab1}), the data values are imposed at the boundary surface and they are used in combination with values in the first and second layers of inner cells to extrapolate the primitive variables into the first and second ghost cell layers, respectively. Since our characteristic BC formulation is in inertial reference frame, the synoptic maps for vector magnetogram and horizontal velocity data are also rotated taking the solar rotation into account. For the rest of the variables, their time variation is determined by solving characteristic compatibility equations (see Eq. (\ref{eq13})) in the ghost cells. These equations involve spatial derivatives of plasma variables which are solved by finite difference approximations taking into account the plasma parameter values in the first and second inner and ghost cell layers. 
\section{Results}
\label{sec3}
\subsection{Relaxation of a dipole magnetic field to a quasi-steady state}
\label{sec3_1}
\begin{figure}[!h]
\centering\hfill\includegraphics[width=0.48\textwidth]{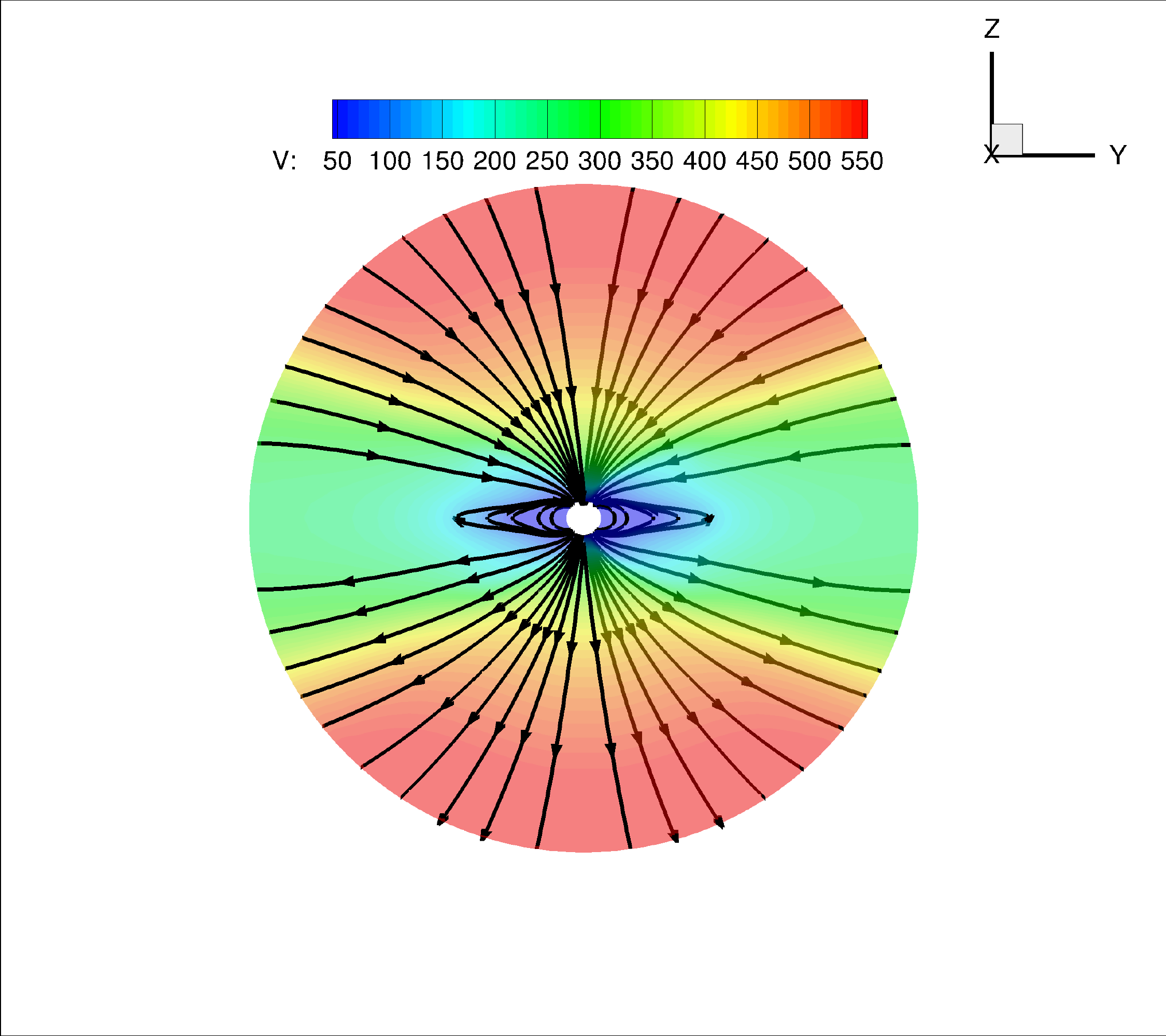}
\includegraphics[width=0.48\textwidth]{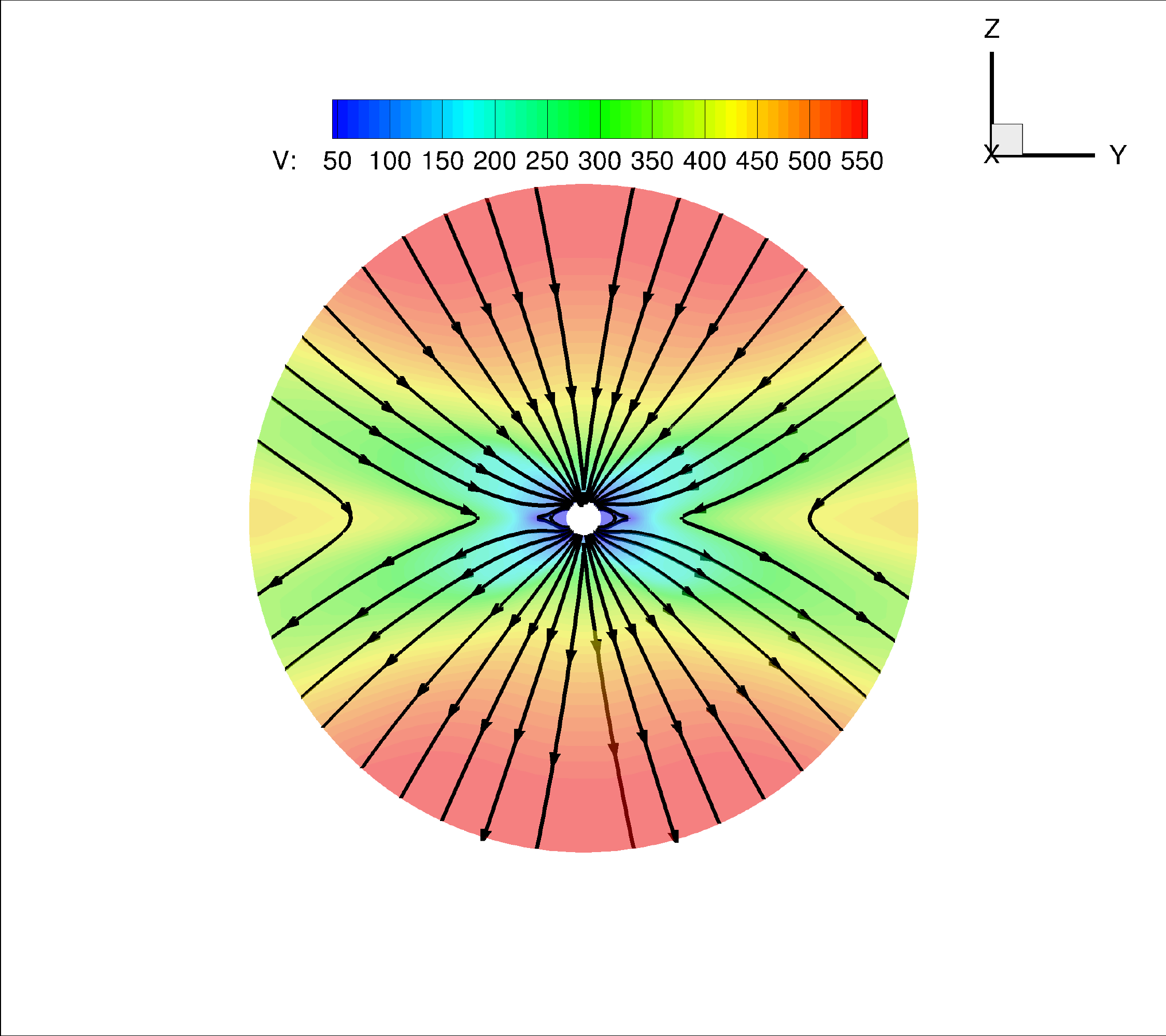}
\begin{center}
\includegraphics[height=5.5cm]{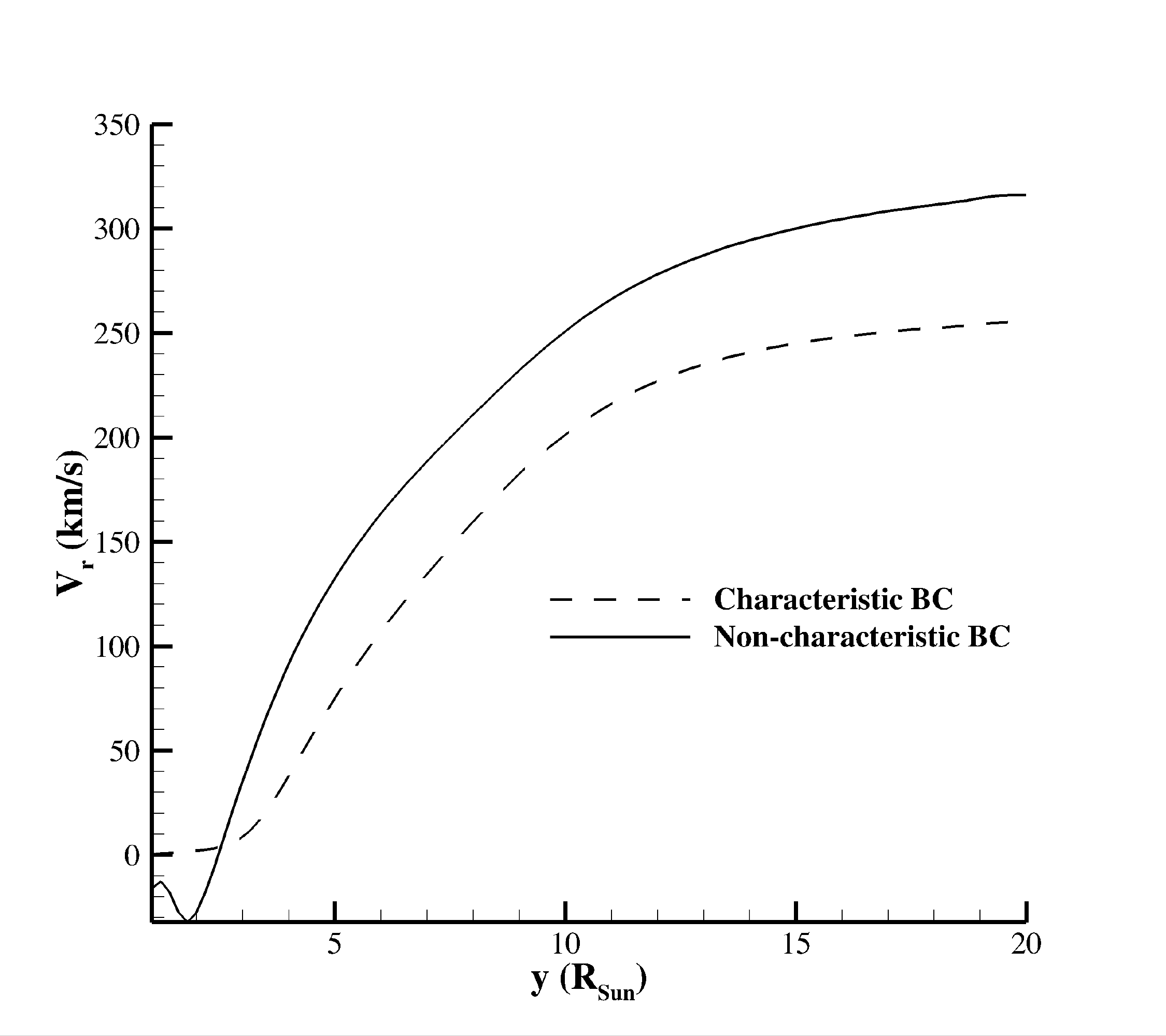}
\end{center} 
\caption{Dipole magnetic field lines relaxed to a quasi-steady state are shown together with the plasma flow speed contours using: (\textit{Left panel}) characteristic BCs and (\textit{Right panel}) one particular choice of non-characteristic BCs, and (\textit{Bottom panel}) radial velocity variations along $y$-axis for characteristic and non-characteristic BCs.}
\label{fig3}
\end{figure}

We present the results of a benchmark test that we performed to validate our characteristic BC formulations. In this test, we impose a dipole magnetic field with magnetic dipole moment equal to the solar dipole moment, $3.5\times 10^{29}$ N m $\textrm{T}^{-1}$, and Parker's isothermal SW model as initial conditions for the magnetic field and hydrodynamic plasma parameters, respectively. Then, we relax this configuration to a quasi-steady state using our characteristic BCs. We also performed another simulation with exactly the same grid, numerical schemes and initial conditions, but this time, using a set of non-characteristic BCs (i.e., $n=1.5\times10^{8}$ $\textrm{cm}^{-3}$,  $B_{r}=(B_{r})_\mathrm{Dipole}$, $B_{\theta}=(B_{\theta})_\mathrm{Dipole}$, $v_{\theta}=v_{\phi}=0$, while $B_{\phi}$, $v_{r}$, and $p$ have zero derivative in the radial direction) instead. Notice that both sets of BCs have the same physical BCs corresponding to the subslow inflow case in Table~\ref{tab1}, whereas the rest of the variables are either obtained from the compatibility equations or their radial derivatives are assumed to be zero. The comparison of the obtained solutions is shown in Figure~\ref{fig3}.

The computational domain has a span of $1.03 R_\odot\le r\le 20 R_\odot$, $0\le\theta\le\pi$, and $0\le\phi\le 2\pi$. The computational grid is 128$\times$56$\times$40 in $r$, $\theta$ and $\phi$ directions, respectively, and is radially stretched. We use a TVD, Roe-type scheme and forward Euler method for spatial and temporal discretizations, respectively. The $\mathbf{\nabla}\cdot\mathbf{B}=0$ constraint is enforced following~\cite{Powell99}. 

In Figure~\ref{fig3}, dipole magnetic field lines that relaxed to a quasi-steady state are shown together with the plasma flow speed contours using characteristic and non-characteristic BCs. Moreover, radial velocity variations are also shown along the $y$-axis for both sets of BCs. 

\begin{figure}[!h]
\centering\hfill\includegraphics[width=0.48\textwidth]{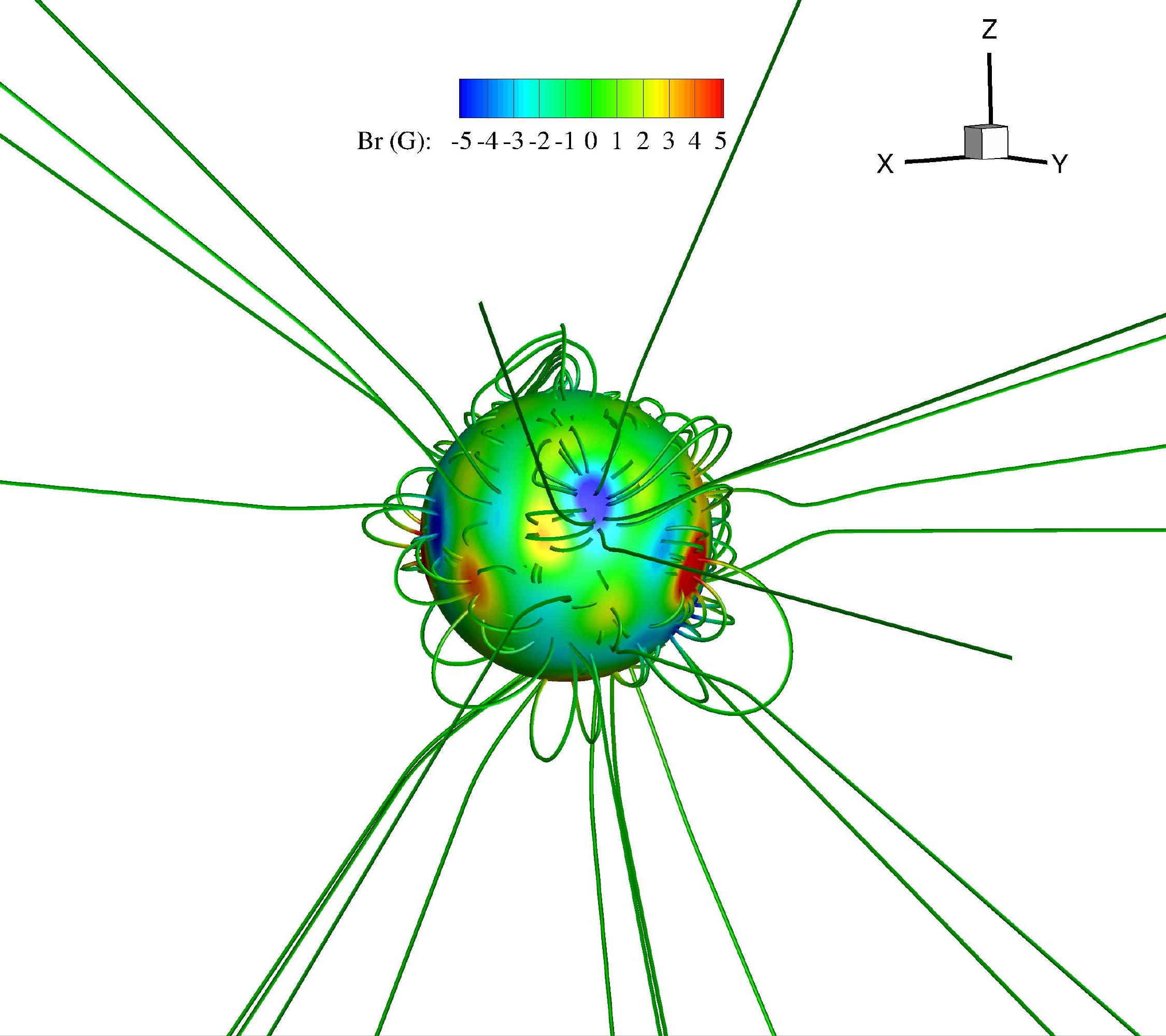}
\includegraphics[width=0.48\textwidth]{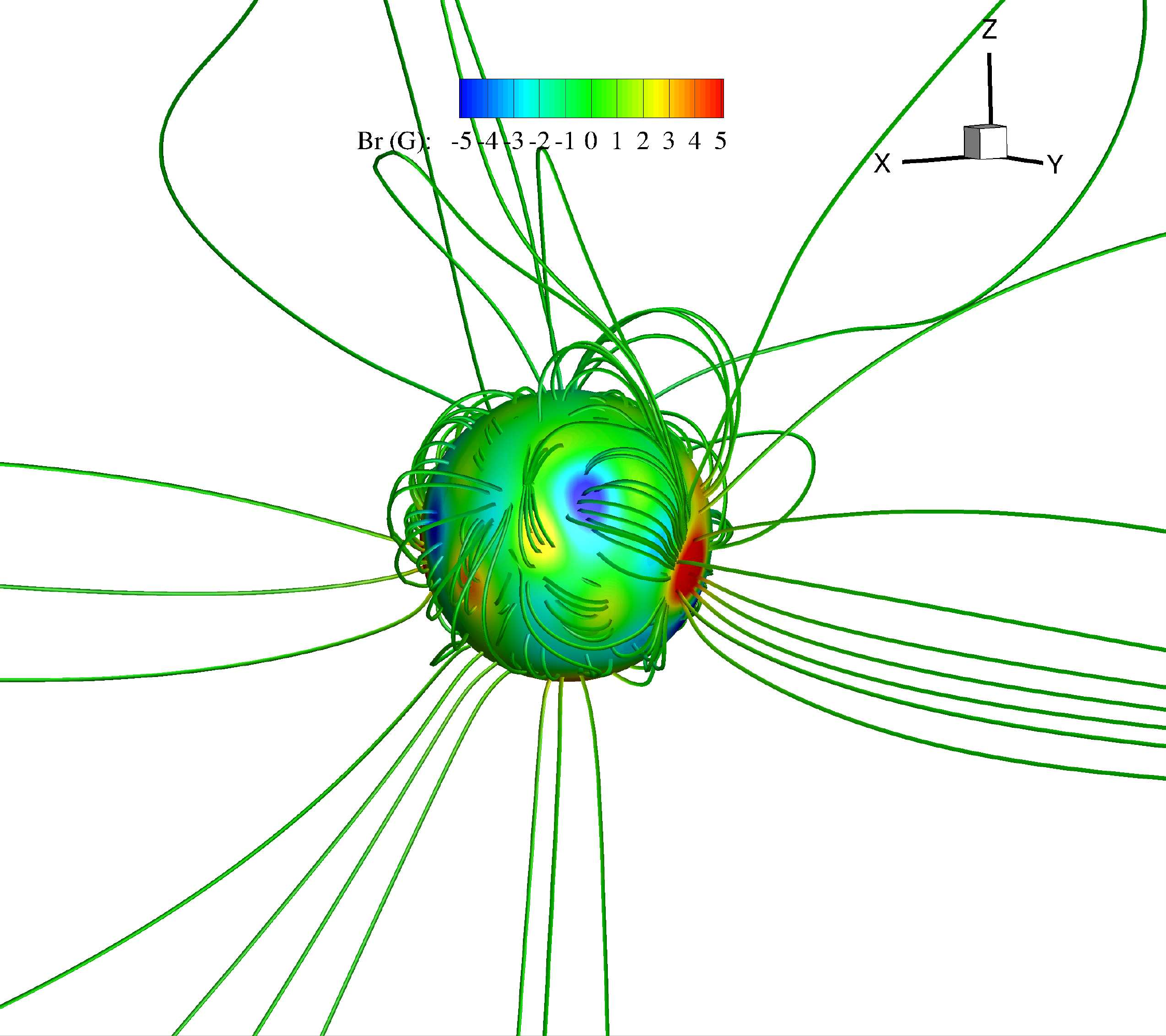}
\begin{center}
\includegraphics[height=5.5cm]{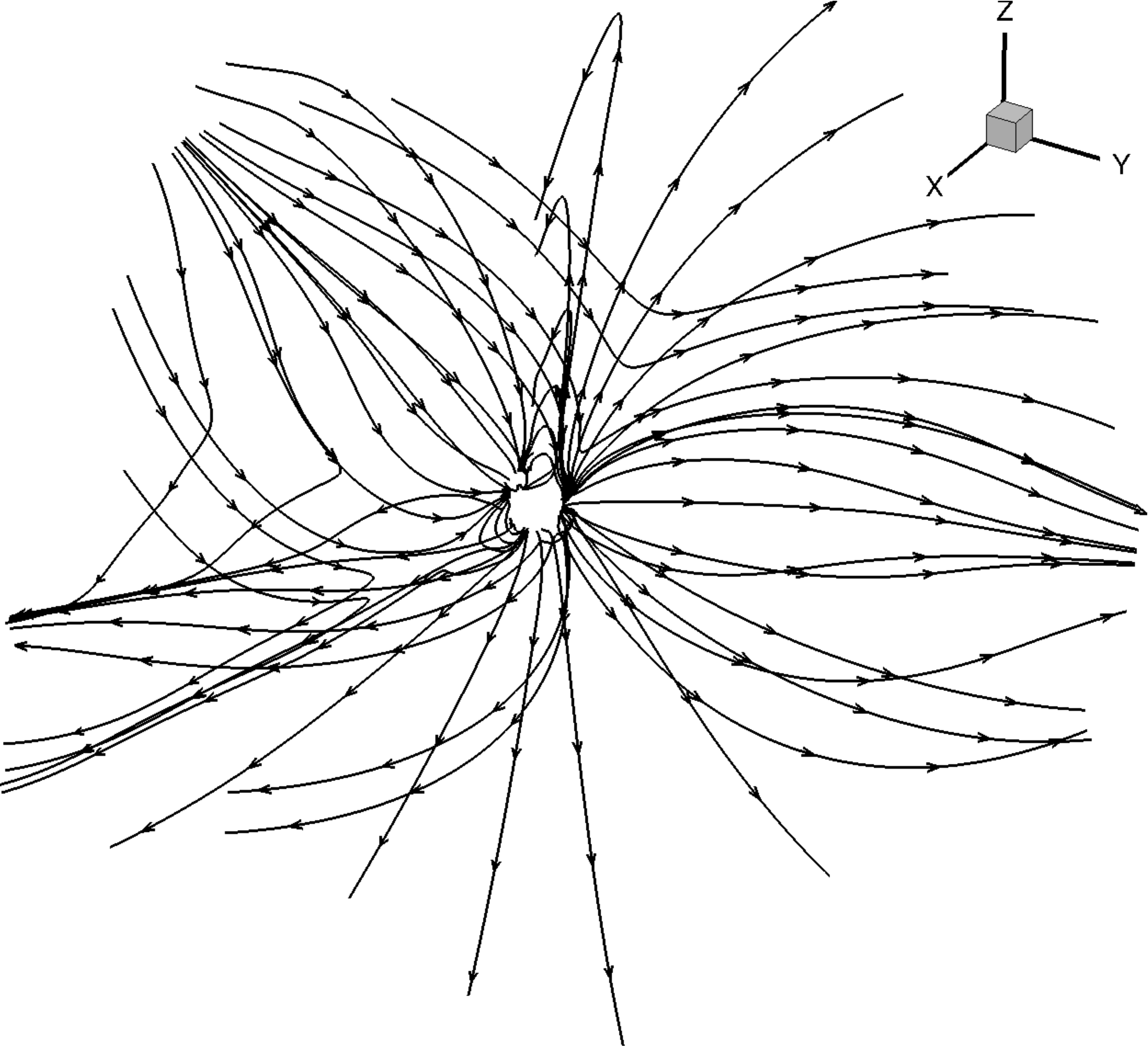}
\end{center} 
\caption{Magnetic field lines in the vicinity of the inner boundary at 1.03$R_{\odot}$ with radial magnetic field line contours for (\textit{Left panel}) initial PFSS solution generated by using the \emph{HMI} LOS magnetogram data observed on 2 January 2014 and (\textit{Right panel}) quasi-steady solution, and (\textit{Bottom panel}) full 3D view of the magnetic field line topology at the quasi-steady state.}
\label{fig4}
\end{figure}

\begin{figure}[!h]
\centering\hfill\includegraphics[width=0.48\textwidth]{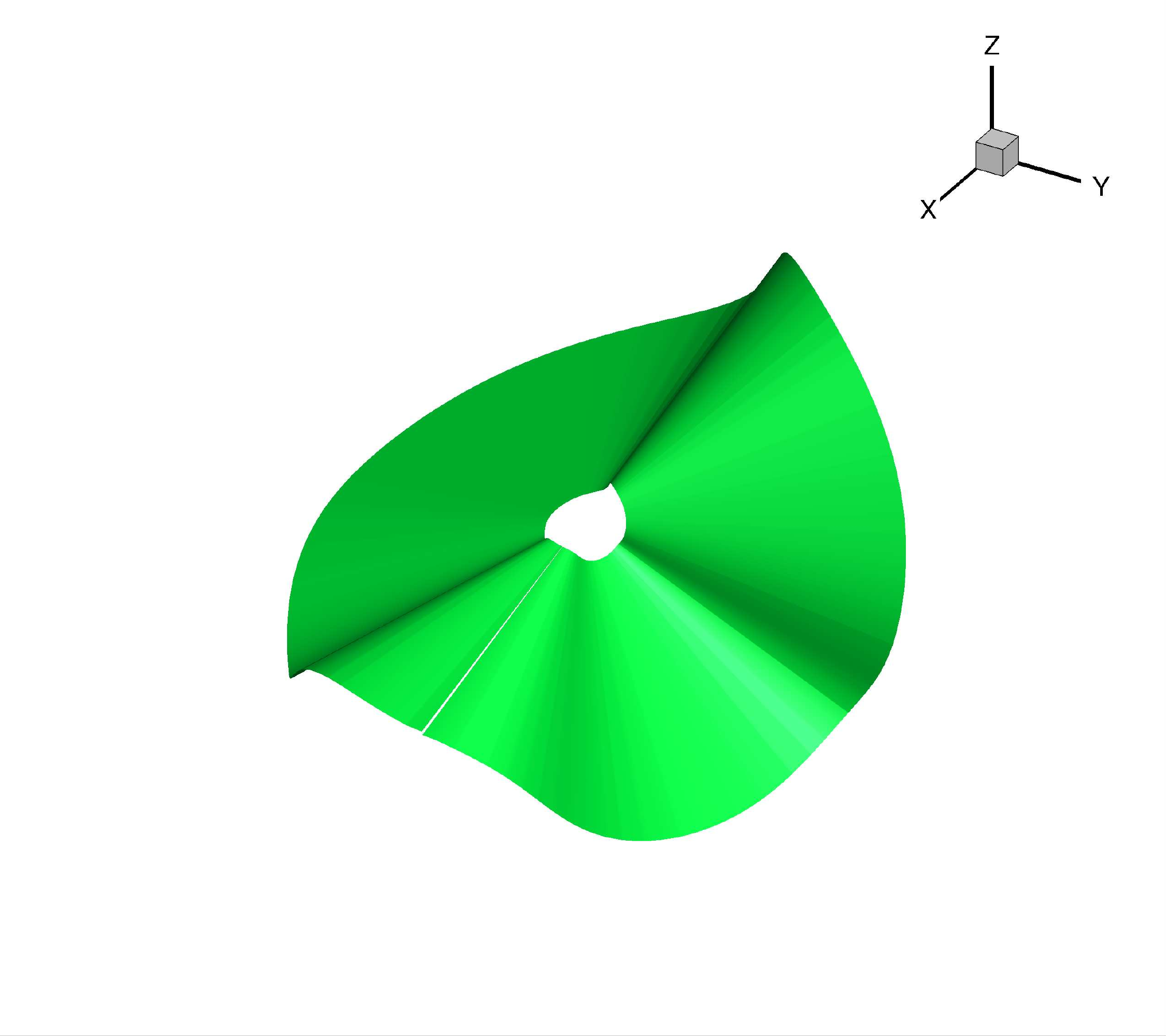}
\includegraphics[width=0.48\textwidth]{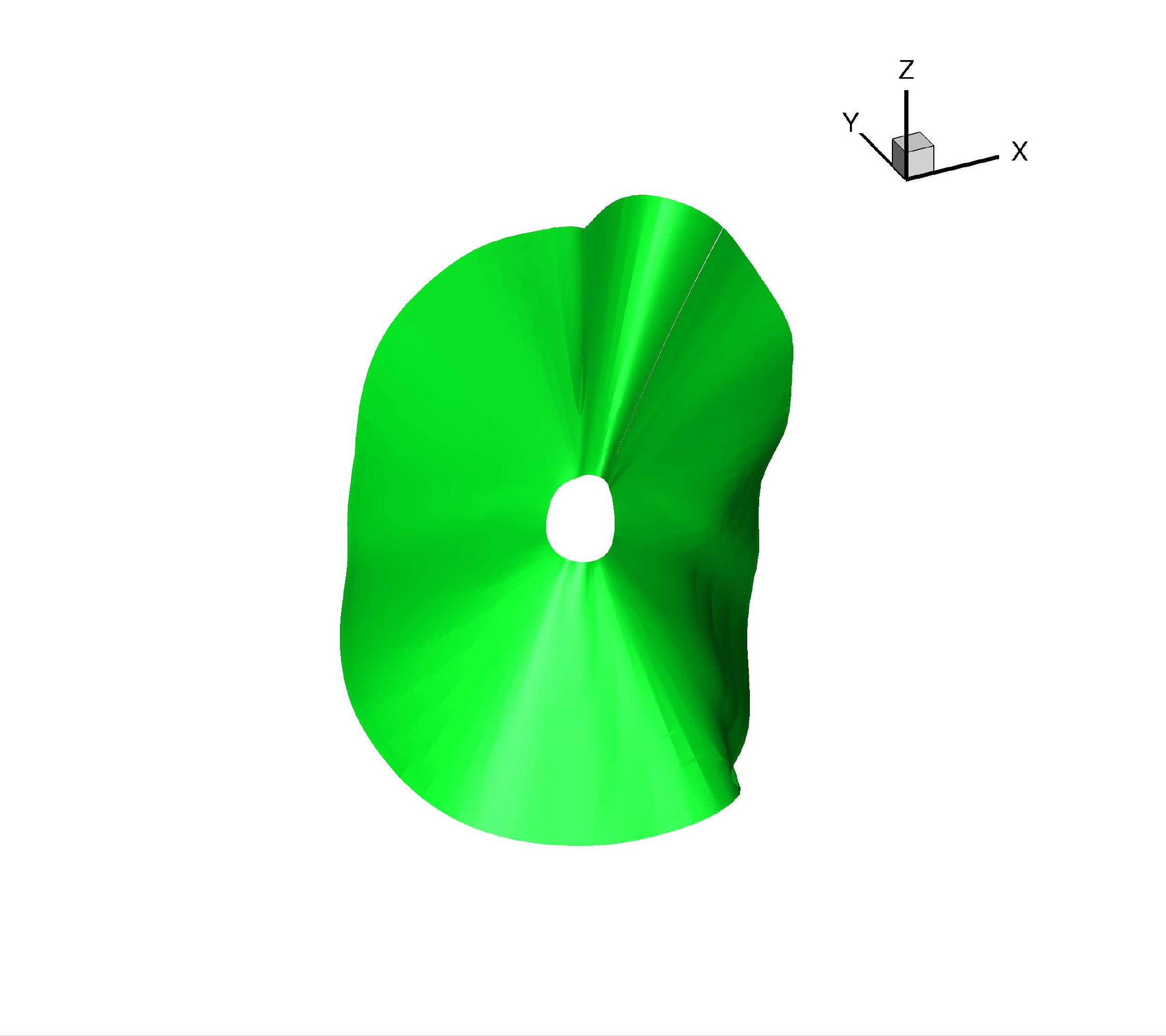}
\caption{Heliospheric current sheet for (\textit{Left panel}) initial PFSS solution generated by using the \emph{HMI} LOS magnetogram data observed on 2 January 2014 and (\textit{Right panel}) quasi-steady solution.}
\label{fig5}
\end{figure}
  
As seen from Figure~\ref{fig3}, using the characteristic BCs leads to a correct magnetic field configuration whereas the non-characteristic approach shows artificial X-points in the vicinity of the Sun. Furthermore, non-characteristic BCs produce a substantial inflow on the solar equatorial plane which does not seem right.
\subsection{Relaxation of the PFSS magnetic field to a quasi-steady state}
\label{sec3_2}
In this section, we present results related to the relaxation of an initial PFSS magnetic field topology generated by using the \emph{HMI} LOS magnetogram data observed on 2 January 2014 to a quasi-steady state. We obtained the PFSS solution by applying the spherical harmonics approach for which we took the spherical harmonics coefficients from the \textit{pfss\_viewer} software in IDL SolarSoft, and using 10 modes. 

The computational domain has a span of $1.03 R_\odot\le r\le 20 R_\odot$, $0\le\theta\le\pi$, and $0\le\phi\le 2\pi$. The computational grid is 128$\times$60$\times$120 in $r$, $\theta$ and $\phi$ directions, respectively, and is radially stretched. The set of non-characteristic BCs applied at the inner boundary is given as follows: $n=1.5\times10^{8}$ $\textrm{cm}^{-3}$, $T=1.3\times 10^{6}$ \textrm{K}, $B_{r}$ is determined from the \emph{HMI} LOS magnetogram data, $v_{\theta}$ and $v_{\phi}$ are the horizontal velocity data shown in Figure~\ref{fig2}, while $B_{\theta}$, $B_{\phi}$, and $v_{r}$  have zero derivative in the radial direction. The horizontal velocity data inferred by the time-distance helioseismology method have 3000$\times$1000 pixels resolution. We applied bilinear interpolation to interpolate the data to our computational grid at the inner boundary in order to be able to utilize in our BCs. It should be noted that differential rotation and meridional flow profiles are removed from the horizontal velocity data. Therefore, we impose empirical formulae for differential rotation~\cite{Komm93a} and meridional flow~\cite{Komm93b} at the entire inner boundary in addition to the data. The simulation is performed in the frame of reference corotating with the Sun.

In Figure~\ref{fig4}, we show the magnetic field line distributions in the vicinity of the inner boundary with radial magnetic field line contours for the initial PFSS and quasi-steady solutions, and a full 3D view of the magnetic field line topology. In Figure~\ref{fig5}, heliospheric current sheets corresponding to the initial PFSS and quasi-steady solutions are presented.   

\section{Conclusions}
\label{sec4}
In this paper, we presented an overview of our data-driven MHD model of SW \textit{based entirely on observations and not relying on any non-reflection principle} in the solar corona. 

For this purpose, we developed a set of genuinely characteristic BCs which gives remarkable results already in a benchmark test.

Our model can be driven by a variety of observational data including \emph{HMI} vector magnetogram and horizontal velocity data, and LOS magnetogram data from \emph{HMI}, \emph{MDI}, \emph{GONG}, and \emph{WSO}. In order to demonstrate this, we presented results related to the simulation of relaxation of a PFSS magnetic field topology to a quasi-steady state which is driven by \emph{HMI} LOS magnetogram and horizontal velocity data using a set of non-characteristic BCs. 

We will present results related to time-dependent simulations of our global corona model in our next publication. 
\ack
The authors acknowledge the support from the NASA grant NNX14AF41G and NSF SHINE grant AGS-1358386. We also acknowledge NSF PRAC award OCI-1144120 and related computer resources from the Blue Waters sustained-petascale computing project. Supercomputer allocations were also provided on SGI Pleiades by NASA High-End Computing Program award SMD-15-5860 and on Stampede by NSF XSEDE project MCA07S033.


\section*{References}
\begin{thebibliography}{29}

\bibitem{Mikic99} Miki\'c Z, Linker J A, Schnack D D, Lionello R and Tarditi A 1999 {\it Phys. Plasmas} {\bf 6} 2217--24
\bibitem{Wu06} Wu S T, Wang A H, Liu Y and Hoeksema J T 2006 {\it Astrophys. J.} {\bf 652} 800--11
\bibitem{Hayashi05} Hayashi K 2005 {\it Astrophys. J. Supp.} {\bf 161} 480--94
\bibitem{vanderHolst10} van der Holst B, Manchester IV W B, Frazin R A, Vasquez A M, Toth G and Gombosi T I 2010 {\it Astrophys. J.} {\bf 725} 1373--83
\bibitem{Yang12} Yang L P, Feng X S, Xiang C Q, Liu Y, Zhao X and Wu S T 2012 {\it J. Geophys. Res - Space Phys.} {\bf 117} A08110
\bibitem{Feng12} Feng X S, Yang L P, Xiang C Q, Liu Y, Zhao X P and Wu S T 2012 {\it Numerical Modeling of Space Plasma Flows} {vol 459} eds N V Pogorelov et al. (ASP Conference Series) p 202
\bibitem{Hirsch94} Hirsch C 1994 {\it Numerical Computation of Internal and External Flows, Volume 2: Computational Methods for Inviscid and Viscous Flows} (John Wiley \& Sons)
\bibitem{Thompson87} Thompson K W 1987 {\it J. Comp. Phys.} {\bf 68} 1--24
\bibitem{Zhao12} Zhao J, Couvidat S, Bogart R S, Parchevsky K V, Birch A C, Duvall T L, Beck J G, Kosovichev A G and Scherrer P H 2012 {\it Sol. Phys.} {\bf 275} 375--90
\bibitem{AN69} Altschuler M D and Newkirk G 1969 {\it Sol. Phys.} {\bf 9} 131--49
\bibitem{Schatten69} Schatten K H, Wilcox J M and Ness N F 1969 {\it Sol. Phys.} {\bf 6} 442--55
\bibitem{Parker58} Parker E N 1958 {\it Astrophys. J.} {\bf 128} 664--76
\bibitem{Nakamizo09} Nakamizo A, Tanaka T, Kubo Y, Kamei S, Shimazu H and Shinagawa H 2009 {\it J. Geophys. Res - Space Phys.} {\bf 114} A07109
\bibitem{Feng10} Feng X, Yang L, Xiang C, Wu S T, Zhou Y and Zhong D 2009 {\it Astrophys. J.} {\bf 723} 300--19
\bibitem{Pogorelov14} Pogorelov N V, Borovikov S N, Heerikhuisen J, Kim T K, Kryukov I A and Zank G P 2014  {\it XSEDE'14 Proc. of the 2014 Annual Conference on Extreme Science and Engineering Discovery Environment} (ACM: New York)
\bibitem{Borovikov09} Borovikov S N, Kryukov I A and Pogorelov N V 2009 {\it Numerical Modeling of Space Plasma Flows} {vol 406} eds N V Pogorelov et al. (ASP Conference Series) p 127
\bibitem{Pogorelov09} Pogorelov N V, Borovikov S N, Florinski V, Heerikhuisen J, Kryukov I A and Zank G P 2009 {\it Numerical Modeling of Space Plasma Flows} {vol 406} eds N V Pogorelov et al. (ASP Conference Series) p 149
\bibitem{Borovikov13} Borovikov S N, Heerikhuisen J and Pogorelov N V 2013 {\it Numerical Modeling of Space Plasma Flows} {vol 474} eds N V Pogorelov et al. (ASP Conference Series) p 219
\bibitem{Pogorelov13} Pogorelov N V, Borovikov S N, Bedford M C, Heerikhuisen J, Kim T K, Kryukov I A and Zank G P 2013 {\it Numerical Modeling of Space Plasma Flows} {vol 474} eds N V Pogorelov et al. (ASP Conference Series) p 165
\bibitem{WS97} Wang Y-M and Sheeley Jr N R 1997 {\it Geophys. Res. Lett.} {\bf 24} 3141--44
\bibitem{Liu17} Liu Y, Hoeksema J T, Sun X and Hayashi K 2017 {\it Sol. Phys.} {\bf 292} 29
\bibitem{Schou12} Schou J et al 2012 {\it Sol. Phys.} {\bf 275} 229--59
\bibitem{Hoeksema14} Hoeksema J T et al 2014 {\it Sol. Phys.} {\bf 289} 3483--3530
\bibitem{Pesnell12} Pesnell W D, Thompson B J and Chamberlin P C 2012 {\it Sol. Phys.} {\bf 275} 3--15
\bibitem{Hoeksema84} Hoeksema J T 1984 {\it Structure and Evolution of the Large Scale Solar and Heliospheric Magnetic Fields} PhD Thesis (Stanford University)
\bibitem{WS92} Wang Y-M and Sheeley Jr N R 1992 {\it Astrophys. J.} {\bf 392} 310--19
\bibitem{SD03} Schrijver C J and DeRosa M L (2003) {\it Sol. Phys.} {\bf 212} 165--200
\bibitem{TVH11} Toth G, van der Holst B and Huang Z 2011 {\it Astrophys. J.} {\bf 732} 102
\bibitem{Thompson90} Thompson K W 1990 {\it J. Comp. Phys.} {\bf 89} 439--61
\bibitem{KPS01} Kulikovskii A G, Pogorelov N V and Semenov A Y 2001 {\it Mathematical Aspects of Numerical Solution of Hyperbolic Systems} (Boca Raton: Chapman \& Hall/CRC Press)
\bibitem{Lionello09} Lionello R, Linker J A and Miki\'c Z 2009 {\it Astrophys. J.} {\bf 690} 902--12
\bibitem{Powell99} Powell K G, Roe P L, Linde T J, Gombosi T I and De Zeeuw D L 1999 {\it J. Comp. Phys.} {\bf 154} 284--309
\bibitem{Komm93a} Komm R W, Howard R F and Harvey J W 1993 {\it Sol. Phys.} {\bf 143} 19--39
\bibitem{Komm93b} Komm R W, Howard R F and Harvey J W 1993 {\it Sol. Phys.} {\bf 147} 207--23
\end{thebibliography}
\end{document}